\documentclass{article}

\usepackage{iftex}
\usepackage{PRIMEarxiv}
\ifPDFTeX
  \usepackage[utf8]{inputenc}
  \usepackage[T1]{fontenc}
\else
  \usepackage{fontspec}
\fi
\usepackage{hyperref}
\usepackage{url}
\usepackage{booktabs}
\usepackage{array}
\usepackage{amsmath,amssymb}
\ifPDFTeX\else
  \usepackage{unicode-math}
\fi
\usepackage[round,authoryear]{natbib}
\usepackage{microtype}
\usepackage{graphicx}
\usepackage{float}
\usepackage{etoolbox}
\usepackage{xcolor}
\usepackage{setspace}
\usepackage{lineno}

\newcommand{\LPD}{\ensuremath{L_{\text{Dice}}}}
\newcommand{\LPJ}{\ensuremath{L_{\text{Jaccard}}}}

\makeatletter
\patchcmd{\thebibliography}{\section*{\refname}}{}{}{}
\makeatother

\hypersetup{hypertexnames=false}

\title{Structural Compression for Phylogenetic Inference under Alignment Instability and Indel-Rich Evolution}
\author{
 \small
 \begin{minipage}[t]{\textwidth}
 \centering
 Zhuoxin Zhang$^{1,2}$,
 Jieyu Wang$^{3,4}$,
 Fengyao Zhai$^{1,5,6}$,
 Jing Wang$^{1,5,6}$,
 Xiaojun Hu$^{1,5,6}$,
 Dayou Zhang$^{2}$,
 Lu Fan$^{7}$,
 Yu Liu$^{1,5}$\thanks{Corresponding author: yu.ernest.liu@bnu.edu.cn} \\
 ~\\
 $^1$Department of Systems Science, Faculty of Arts and Sciences, Beijing Normal University, Zhuhai 519087, China.\\
 $^2$College of Forestry and Landscape Architecture, South China Agricultural University, Guangzhou 510640, China.\\
 $^3$Guangdong Provincial Genomics Data Center, BGI Research, Shenzhen 518120, China.\\
 $^4$BGI Research, Shenzhen 518083, China.\\
 $^5$International Academic Center of Complex Systems, Beijing Normal University, Zhuhai 519087, China.\\
 $^6$School of Systems Science, Beijing Normal University, Beijing 100875, China.\\
 $^7$Department of Ocean Science and Engineering, Southern University of Science and Technology (SUSTech), Shenzhen 518055, China.
 \end{minipage}
}
\date{}

\begin{document}
\maketitle
%\linenumbers

\begin{abstract}
Phylogenetic inference traditionally relies on aligned characters under substitution models, but this framework becomes less reliable when alignments are unstable or when evolution is dominated by insertions, deletions, repeats, and other structural changes. We adapt Ladderpath as an alignment-free distance approach for phylogenetic inference. Motivated by algorithmic information theory, Ladderpath decomposes sequences into derived, reusable units (``ladderons'', rather than fixed-length $k$-mers) organized hierarchically, from which pairwise distances are computed. The premise is that shared derived sequence structure, including repeated or reused segments that are poorly represented by column-wise substitutions, can retain phylogenetic information. The bacteriophage T7 known lineage, the cpSSR repeat-rich marker, and a cytochrome~$c$ protein dataset confirm that Ladderpath recovers topologies consistent with the known experimental history or with established alignment-based methods. Its advantage emerges under stress: in block-translocation and indel-dominated simulations Ladderpath remains stable while alignment-dependent pipelines deteriorate; on banana mitochondrial and plastome genomes it scales to genome length and captures the expected contrast between organellar histories, all from unaligned input. These results support Ladderpath as an alignment-free, structurally informed method that could complement standard pipelines in cases where higher-order sequence structure carries phylogenetic signal.
\end{abstract}


%\newpage
%%%%%%%%%%%%%%%%%%%%%%%%%
%%%%%%%%%%%%%%%%%%%%%%%%%
\section{Introduction}

Improving phylogenetic reconstruction remains a central task in evolutionary biology. Most mainstream frameworks, including maximum likelihood (ML), Bayesian inference (BI), and maximum parsimony (MP), infer history primarily from aligned characters under site-wise substitution models \citep{yang2006}. They are highly effective when homology is clear and sequence change is mainly substitutional. Their limits become more visible, however, when alignment itself becomes unstable or when sequence evolution is shaped substantially by insertions, deletions, repeat expansion, duplication, recombination, inversion, and larger-scale rearrangement. These processes can carry genuine phylogenetic information, yet they are not naturally represented in a framework centered on aligned site states.

This problem has motivated a broad literature on alignment-free sequence comparison and phylogenetics \citep{haubold2014, zielezinski2017}. Existing approaches include $k$-mer-based summaries \citep{zielezinski2017}, Composition Vector Trees (CVTree) \citep{zuo2015}, MinHash-style sketching \citep{ondov2016}, and more recent Lempel-Ziv-related approaches such as Vclust \citep{zielezinski2025}. Earlier global-sequence approaches also explored phylogenetic signal outside strict site-wise comparison \citep{chapus2005}. Collectively, these methods reduce the dependence on multiple-sequence alignment and often offer major gains in speed, scalability, or robustness. They are valuable when alignment quality or computational cost is the main bottleneck. At the same time, most still encode sequences mainly through composition counts, short-word matches, or flattened complexity summaries rather than through explicitly nested structural reuse.

That distinction matters if biological sequences are viewed not merely as strings of independent sites, but as products of evolutionary tinkering \citep{jacob1977}. Duplication, repetition, rearrangement, indel accumulation, and modular reuse can generate hierarchical organization within sequences. Such reuse is consistent with the central role of gene duplication and divergence in molecular evolution \citep{ohno1970evolution} and with broader views of biological modularity across molecular systems \citep{hartwell1999modular}. Recent Ladderpath-based work further suggests that nested reusable organization can be measured directly in biological sequences \citep{Liu2021SciAdv, liu2024PRR}. From this perspective, structural reuse is not simply noise around a substitutional core. It may itself preserve part of the historical signal. A method that measures this organization directly could therefore complement standard phylogenetic inference, especially in regimes where alignment quality deteriorates or where indel-driven change contributes substantially to divergence.

Related ideas have appeared in phylogenetics from several directions. Rare genomic changes, including indels, retroposon insertions, gene-order changes, and duplications, have long been discussed as phylogenetic markers because they can represent derived historical events rather than independent site substitutions \citep{rokas2000rare}. Gap patterns and variable regions can also contain tree signal that is lost when they are discarded or treated only as alignment uncertainty \citep{dessimoz2010gaps}, while indel-aware alignment--phylogeny models treat insertion, deletion, and alignment uncertainty as part of the evolutionary process itself \citep{thorne1991evolutionary, redelings2005joint}. In parallel, whole-sequence information-distance approaches show that phylogenetic signal can be recovered without first reducing sequences to fixed aligned columns \citep{otu2003new}. These precedents motivate the question addressed here: if sequence history leaves traces in reusable higher-order modules, can those modules be measured directly and used as phylogenetic signal?

\emph{Ladderpath} fits this problem because it was developed to measure sequence complexity through recursive reuse of substructures \citep{liu2022}, not originally as a phylogenetic method. Its central idea is to decompose a sequence into reusable units, termed \textit{ladderons}, and to describe the sequence through modular assembly rather than through aligned substitutions alone \citep{Liu2024npj, Liu2024PepHiRe, Liu2025Alg}. Here we adapt Ladderpath to phylogenetic inference for the first time by deriving distances from ladderon-based decompositions. The test is practical: whether this structural representation recovers useful signal when standard alignment-based pipelines, or flatter alignment-free summaries, become less reliable.

We ask where Ladderpath is informative for phylogenetic inference, and where it is not. The study combines a known experimental phylogeny in bacteriophage T7 \citep{hillis1992, hillis1994}, simulations under standard, alignment-difficult, and indel-dominated conditions, and empirical datasets selected for different forms of structural signal. These include a repeat-rich chloroplast simple sequence repeat (cpSSR) marker \citep{provan2001}, banana organellar genomes with contrasting inheritance histories \citep{wu2021banana}, and the classic Fitch \& Margoliash cytochrome~$c$ proteins \citep{fitch1967}. A Brassicaceae collinearity analysis is retained as an extended structural application in the Supplementary Information (SI). The comparator panel comprises three alignment-based methods (ML, MP, BI), two classical distance methods (NJ, UPGMA), and one alignment-free composition method (CVTree); their specific roles and the relationship to Ladderpath are detailed in Methods. The aim is not to claim a universal advantage, but to define the conditions under which Ladderpath is a useful alignment-free structural complement.

%%%%%%%%%%%%%%%%%%%%%%%%%
%%%%%%%%%%%%%%%%%%%%%%%%%
\section{The Ladderpath framework for phylogenetic inference}

\subsection{Recap of the ladderpath approach}

Ladderpath was introduced previously as an algorithmic-information-theoretic framework for quantifying sequence complexity through hierarchical reuse \citep{liu2022, liu2024PRR}. Here we only recap the parts needed for phylogenetic use. The key operation is reconstructing a target sequence from primitive symbols while allowing previously generated substructures to be reused, an assumption that parallels evolutionary tinkering \citep{jacob1977}.

Starting from a basic set $S_0$ of primitive symbols, a \emph{generation operation} combines existing blocks into a new block and places it at a higher level of a partially ordered multiset. Each operation corresponds to one reconstruction step. For a target sequence $X$, the \emph{ladderpath} is the shortest reconstruction route from $S_0$ under this reuse rule; the intermediate reusable blocks are called \emph{ladderons}. The length of this shortest route is the \emph{ladderpath index}, denoted by $\lambda(X)$.

Two auxiliary quantities are useful for interpreting this index. The \emph{size index} $S(X)$ is the length of the shortest trivial reconstruction with no reuse, equal to the sequence length for character strings. The \emph{order index}, $\omega(X) := S(X) - \lambda(X)$, measures how many reconstruction steps are saved by reusing internal structure. Thus, $\lambda$ captures the irreducible reconstruction cost, whereas $\omega$ captures the reusable component.

\begin{figure}[!htp]
  \centering
  \includegraphics[width=0.8\linewidth]{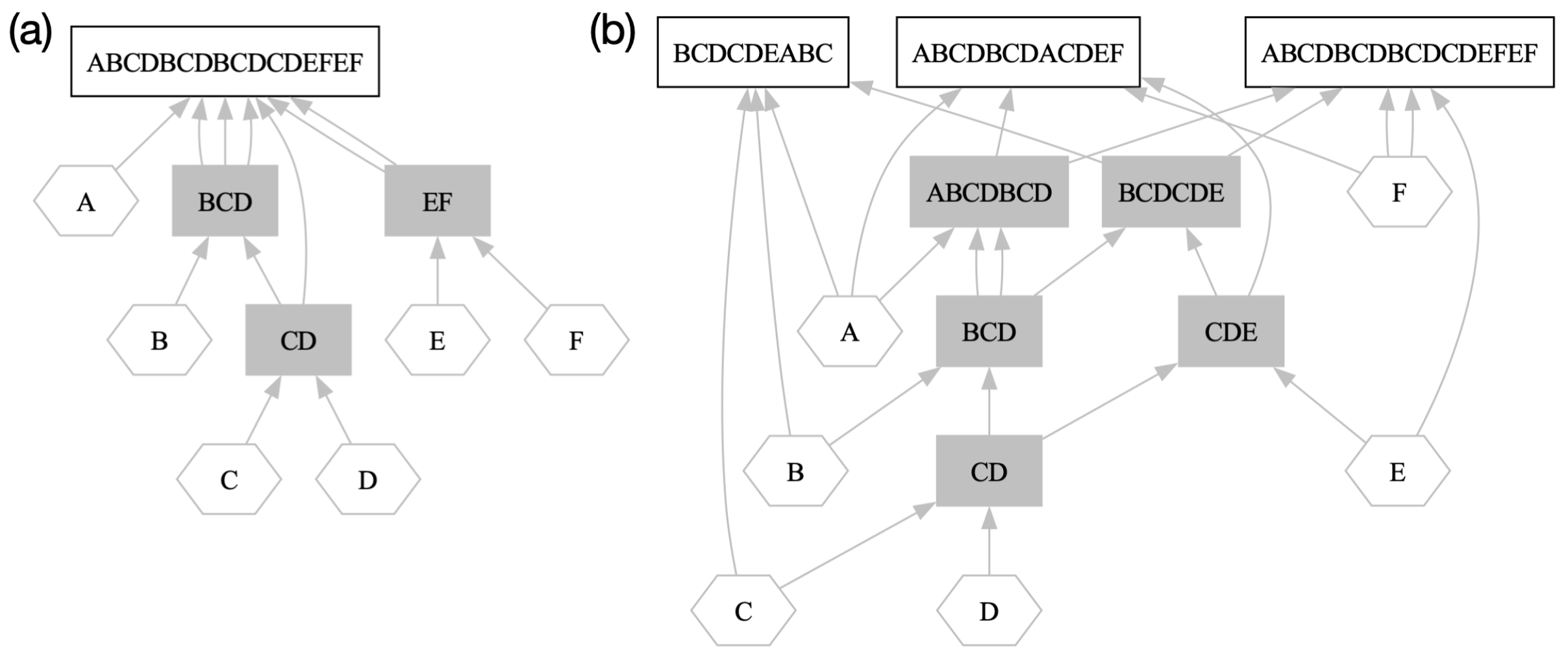}
  \caption{Ladderpath representation. (a) Single sequence decomposed into a laddergraph with reused substrings as higher-order ladderons. (b) Multiple-sequence laddergraphs expose shared and sequence-specific ladderons used to define pairwise distances.}
  \label{fig:ladderpath-theory}
\end{figure}

A \emph{ladderpath} can be represented as a \emph{laddergraph}, a directed acyclic graph in which primitive symbols and higher-order ladderons are linked by composition relationships. Fig.~\ref{fig:ladderpath-theory}a shows a single-sequence example, where the target sequence is reconstructed through repeated modules such as CD, BCD, and EF. Fig.~\ref{fig:ladderpath-theory}b shows the multiple-sequence case; shared and sequence-specific ladderons are the structural quantities used below to define distances.

%\newpage
%%%%%%%%%%%%%%%%%%%%%%%%%
\subsection{The Ladderpath-Dice distance for phylogenetic inference}

The ladderpath representation functions as a compression-like description: repeated substructures are constructed once and then reused. To compare two sequences $X$ and $Y$, we locate the two target nodes in the same laddergraph and evaluate the reconstruction steps required to contain both targets. Shared ladderons are counted once in this reconstruction graph, so the resulting index reflects shared internal structure.

The main distance used in this study is the Ladderpath-Dice distance, defined previously in \citet{Liu2026distance}:
\begin{equation}
  L_{\text{Dice}}(X, Y) = \frac{\lambda'(X, Y) - \dfrac{\lambda'(X) + \lambda'(Y)}{2}}{\dfrac{\lambda'(X) + \lambda'(Y)}{2}}
  \label{eq:l-dice}
\end{equation}
Here, $\lambda'(X):=\lambda(X)-1$ is the adjusted ladderpath index for a single target, and $\lambda'(X,Y):=\lambda(X,Y)-2$ is the adjusted ladderpath index for the reconstruction graph containing both targets. This small adjustment---subtracting 1 or 2---accounts for extracting the final target string or strings during reconstruction; more generally, if $n$ targets are extracted, then $n$ is subtracted. This distance is 0 when $X=Y$ and approaches 1 when the two sequences share no reusable substructures.

\LPD~is a semi-metric: it satisfies non-negativity, identity of indiscernibles, and symmetry, but not the triangle inequality \citep{Liu2026distance}. A closely related \LPJ version provides the corresponding metric check and is reported in SI Section S1. Dice and Jaccard give the same qualitative conclusions in the present analyses. We use Dice in the main text because Dice-type overlap measures are widely used for sparse shared-structure data and because \LPD~gives slightly better empirical phylogenetic accuracy across most conditions. The resulting pairwise distance matrix can be supplied to standard distance-based tree builders; the main text reports \LPD with neighbor joining (NJ) and/or the unweighted pair group method with arithmetic mean (UPGMA) where relevant, while the \LPJ counterparts are collected in the SI.

%%%%%%%%%%%%%%%%%%%%%%%%%
\subsection{Toy model: why structural reuse can affect topology}

This toy model is a conceptual example rather than a statistical benchmark. It shows a case in which evolutionary history is carried by reuse and reassembly of sequence modules. In a duplication-rich history, a descendant may preserve much of the same alphabetic content as its ancestor while changing how those letters are assembled into larger reusable blocks. A strictly site-based or flat-composition comparison can therefore miss part of the historical signal, whereas a laddergraph-based representation can preserve the reuse history explicitly.

\begin{figure}[!htp]
  \centering
  \includegraphics[width=1.0\linewidth]{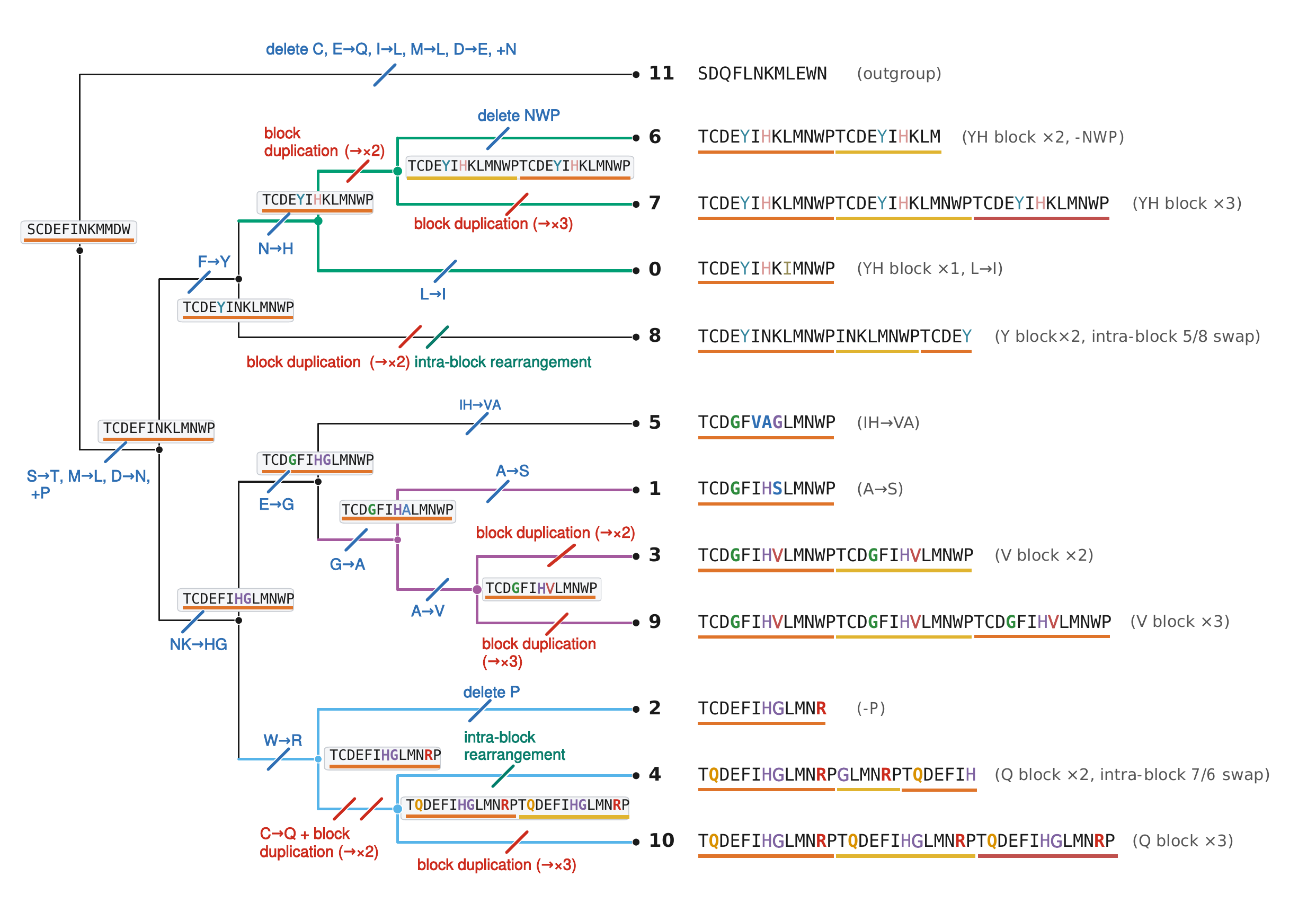}
  \caption{Designed construction history for the 12-taxon toy protein dataset, used as the reference topology in Fig.~\ref{fig:toy-comparison}. Starting from a common ancestral sequence (left), each branch applies labeled edits---point substitutions and indels (blue), block duplications producing tandem repeats (red), and intra-block rearrangements (green)---to yield the 12 extant sequences (right; taxon 11 is the outgroup). Coloured underlines mark blocks.}
  \label{fig:toy-ground-truth}
\end{figure}

Figure~\ref{fig:toy-ground-truth} shows the designed construction history used as the reference topology. Starting from a common ancestral sequence, the 12 taxa were produced by combining point substitutions, indels, block duplications (generating two- and three-fold tandem repeats), and intra-block rearrangements, so that several lineages are defined primarily by how shared modules are duplicated and reassembled rather than by isolated site substitutions. For the cross-method comparison, ML, MP, BI, NJ, and UPGMA were run from the same MAFFT alignment of the 12-taxon protein dataset (sequence lengths 12--39 residues; 39 alignment columns), while the alignment-free comparators (CVTree and the ladderpath reconstructions \LPD+NJ and \LPD+UPGMA) were computed directly from the raw sequences. All inferred trees and the reference topology were rerooted on the designated outgroup before computing rooted nRF.

\newpage

\begin{figure}[!htp]
  \centering
  \includegraphics[width=1.0\linewidth]{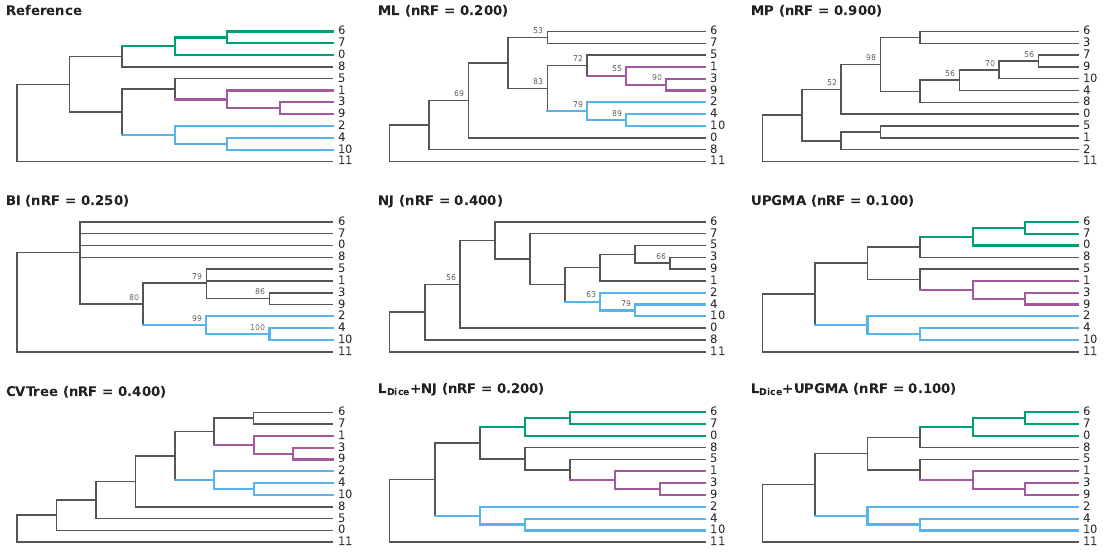}
  \caption{Tree comparison for the toy protein dataset. Top-left: designed reference; other panels: inferred trees per method. Reference clades color-coded (\{1,3,9\} purple, \{2,4,10\} blue, \{0,6,7\} green); matching clades share the same color. Rooted nRF annotated per panel. Branch labels: UFBoot (ML), non-parametric BS $B=1000$ (MP, NJ), PP$\times 100$ (BI).}
  \label{fig:toy-comparison}
\end{figure}

Figure~\ref{fig:toy-comparison} compares the resulting trees against the designed reference topology. By rooted nRF, UPGMA and \LPD+UPGMA are both lowest at $0.100$, followed by ML and \LPD+NJ at $0.200$, BI at $0.250$ (this value reflects an unresolved polytomy in the BI consensus tree that lowers its maximum possible RF), NJ and CVTree at $0.400$, and MP highest at $0.900$. The qualitative signal is sharper in the color-coded clades in Fig.~\ref{fig:toy-comparison}: three reconstructions---\LPD+UPGMA, \LPD+NJ, and classical UPGMA---recover all three designed reference clades (\{1,3,9\}, \{2,4,10\}, and \{0,6,7\}). ML and CVTree each recover two (\{1,3,9\} and \{2,4,10\}); NJ and BI each recover only \{2,4,10\}; MP recovers none. Notably, both Ladderpath reconstructions recover the full set of module-defined clades directly from the raw sequences, matching the alignment-based method that scored best on this dataset (classical UPGMA) and outperforming the other alignment-free comparator (CVTree).

Each Ladderpath reconstruction has a classical counterpart that uses the same tree-builder but a model-based substitution distance: \LPD+NJ versus NJ, and \LPD+UPGMA versus UPGMA. Comparing within a pair therefore isolates the effect of switching to the \LPD distance, so any difference in the recovered topology reflects that change alone. Under NJ, the substitution distance recovers only one clade while the Ladderpath distance recovers all three; under UPGMA, both recover all three. On this designed dataset, the improvement under NJ is attributable to the \LPD distance rather than the tree-builder---a proof of concept that, for this particular reuse-driven history, the structural distance can itself carry signal recoverable without alignment.

%\newpage
%%%%%%%%%%%%%%%%%%%%%%%%%
%%%%%%%%%%%%%%%%%%%%%%%%%
\section{Results}

The Results first use a known experimental phylogeny as an anchor, then test Ladderpath under standard, alignment-difficult, and indel-dominated simulations, and finally examine representative empirical datasets. Tree agreement is summarized by the normalized Robinson--Foulds distance (nRF), where lower values indicate closer topological agreement. Additional supporting comparisons are provided in SI Sections S2 and S3.

%%%%%%%%%%%%%%%%%%%%%%%%%
\subsection{Anchor validation on a known experimental phylogeny}

The experimental bacteriophage T7 lineage provides an anchor because its branching history is known independently from the sequence data. Hillis and colleagues generated this phylogeny through controlled serial passage with predefined lineage splitting, so the reference tree was established by experimental design rather than inferred retrospectively from molecular characters \citep{hillis1992, hillis1994}. T7 is therefore a calibration case, not a broad empirical validation panel.

Table~\ref{tab:t7-nrf} summarizes the nRF distance between each inferred tree and the experimentally established lineage. All eight reconstructions recover the three top-level reference clades \{J,M\}, \{K,L\}, and \{N,O,P,Q\}, but they differ on the placement of M within \{J,M\} and on the \{O,P\} sub-clade. \LPD+UPGMA is the only method that recovers the full reference topology (nRF = 0.000). ML, MP, NJ, and BI retain \{J,M\} but mis-resolve \{O,P\} (nRF = 0.143; BI's 0.077 reflects a polytomy in the consensus). UPGMA, CVTree, and \LPD+NJ retain \{O,P\} but mis-place M outside \{J,M\} (nRF = 0.143). The corresponding \LPJ-based result is reported in SI Section S3.

\begin{table}[!ht]
\centering
\small
\resizebox{0.75\textwidth}{!}{%
\begin{tabular}{lcccccccc}
\toprule
	 & ML & MP & BI & NJ & UPGMA & CVTree & \LPD+NJ & \LPD+UPGMA \\
\midrule
nRF & 0.143 & 0.143 & 0.077 & 0.143 & 0.143 & 0.143 & 0.143 & 0 \\
\bottomrule
\end{tabular}}
\vspace{0.6em}
\caption{Rooted nRF against the known experimental T7 lineage; lower values indicate closer agreement.}
\label{tab:t7-nrf}
\end{table}

\begin{figure}[!ht]
  \centering
  \includegraphics[width=1.0\linewidth]{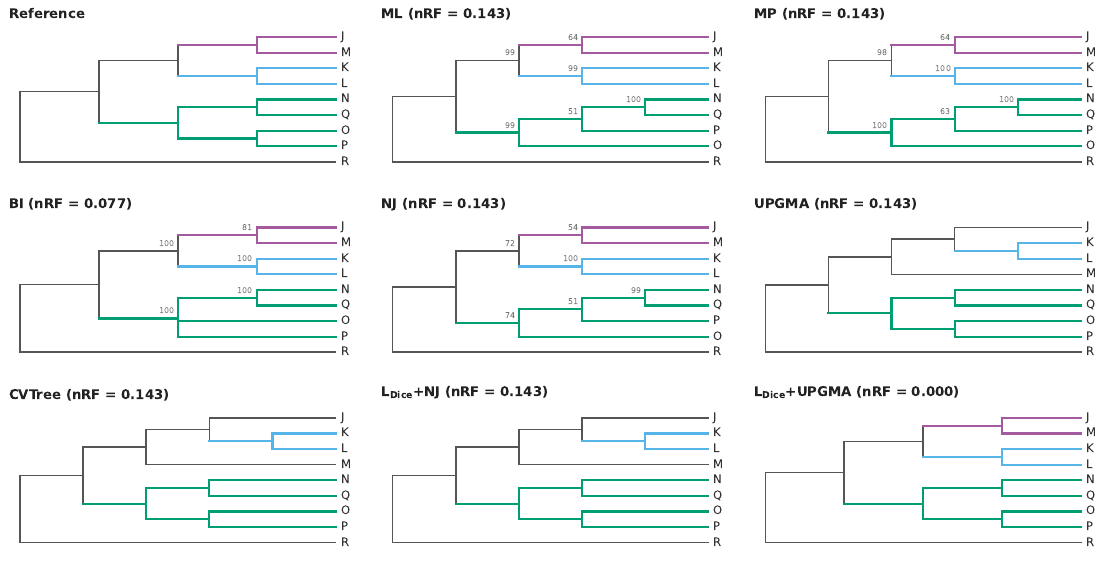}
  \caption{Tree comparison for the T7 experimental lineage. Top-left: reference; other panels: inferred trees per method. Reference clades color-coded (\{J,M\} purple, \{K,L\} blue, \{N,O,P,Q\} green); matching clades share the same color. Rooted nRF annotated per panel. Branch labels: UFBoot (ML), non-parametric BS $B=1000$ (MP, NJ), PP$\times 100$ (BI).}
  \label{fig:t7}
\end{figure}

%\newpage
T7 alone does not establish broad empirical validity. It shows that a structural representation can closely reproduce a known, non-simulated evolutionary history. The remaining sections test when the same representation remains useful under standard, alignment-difficult, and structurally dominated conditions.

%%%%%%%%%%%%%%%%%%%%%%%%%
%%%%%%%%%%%%%%%%%%%%%%%%%
\subsection{Simulation benchmarks across application regimes}

The simulation benchmarks separate three regimes: standard divergence under reliable alignment, block-translocation stress on alignment, and indel-dominated evolution with weak substitutional signal.

%%%%%%%%%%%%%%%%%%%%%%%%%
\subsubsection{Baseline performance under standard conditions}
\label{sec:baseline}

We first establish the standard baseline. This benchmark asks how Ladderpath behaves when alignment is reliable and sequence evolution remains largely substitution-centered.

The divergence-gradient benchmark used 20 taxa, 4,000 nt root sequences, 30 replicates per condition, GTR+$\Gamma$ substitutions, light indel noise, and simple sequence repeats. The manipulated parameter was overall tree scale, a branch-length multiplier controlling expected sequence divergence. Five settings were used (A1--A5: 0.02, 0.06, 0.15, 0.30, and 0.50). Mean pairwise distances (0.027--0.278), MAFFT alignment gap fractions (0.004--0.081), and alignment lengths are diagnostics rather than independently varied parameters and are reported in SI Section S2. MAFFT alignment quality remained reliable throughout the gradient.

\begin{figure}[!htp]
  \centering
  \includegraphics[width=1.0\linewidth]{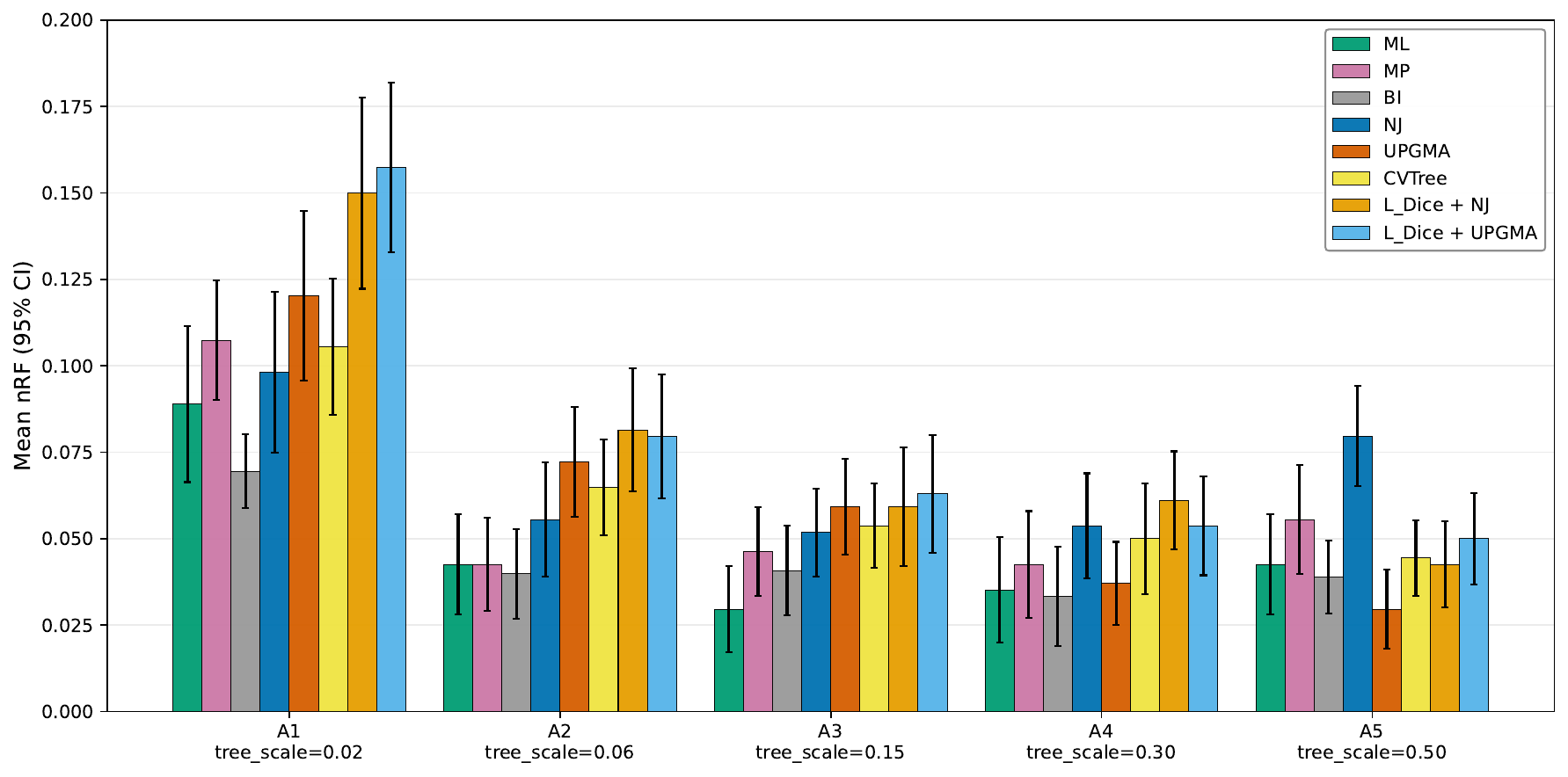}
  \caption{Standard divergence-gradient baseline (A1--A5). Under alignment-reliable, substitution-centered conditions, \LPD+NJ is comparable to established methods across moderate-to-high divergence but not the strongest at very low divergence. Bars show mean rooted nRF across 30 replicates; error bars are 95\% confidence intervals.}
  \label{fig:divergence-gradient-trend}
\end{figure}

Figure~\ref{fig:divergence-gradient-trend} summarizes the nRF trend. This is the easiest of the three simulation benchmarks: across all methods and all five divergence levels, mean nRF remains below 0.16. \LPD+NJ stays within the same broad accuracy range as the comparator methods, with mean nRF decreasing from 0.150 in A1 to 0.043 in A5. The main exception is shallow divergence. In A1, \LPD+NJ has higher error than ML (0.150 vs 0.089; Holm-adjusted $p=0.007$) and BI (0.150 vs 0.069; $p=10^{-4}$). In A2, ML and BI remain lower than \LPD+NJ (0.081 vs 0.043 and 0.040; $p=0.004$ and $p=0.005$). Here and below, Holm-adjusted $p$ values correct for multiple method comparisons within each condition. By A3--A5 the differences between \LPD+NJ and the model-based methods are no longer significant after Holm correction.

Among alignment-free methods, the baseline differences are small. Within the Dice-based Ladderpath results, NJ and UPGMA produce no statistically detectable differences in this benchmark (Holm-adjusted comparisons non-significant; Cliff's $\delta$ effect sizes $|\delta|\leq 0.07$), and the corresponding \LPJ~checks are reported in SI Section S2. \LPD+NJ is also statistically indistinguishable from CVTree across the gradient.

The baseline result is simple: under standard, alignment-reliable conditions, the ladderpath approach is viable but not dominant. It becomes comparable to the standard methods once divergence is no longer extremely shallow, but ML and BI remain stronger in low-divergence datasets where substitutions are sparse and alignment is clean; MP also has an advantage in some shallow-divergence settings. This baseline sets up the next two benchmarks, where alignment reliability or substitutional signal is deliberately weakened.

%%%%%%%%%%%%%%%%%%%%%%%%%
\subsubsection{Robustness when alignment becomes unreliable}

Having established the standard baseline, we next ask whether the ladderpath approach remains usable when alignment becomes the weak point. We used block translocation as the stressor because it preserves much of the sequence material while changing block order. This directly challenges positional homology in multiple-sequence alignment without increasing the substitution rate or the background indel rate.

The manipulated axis was block-order disruption. Six conditions (B1--B6) increased the expected number of block-translocation events per branch from 0 to 20; stronger conditions also used larger translocated blocks. The substitution model and light background indel process were held fixed. Each condition contained 30 simulated 30-taxon datasets of 1{,}500 nt. Core parameter values are given in SI Section S2. This benchmark is therefore distinct from the indel-dominated benchmark below: here the point is not to add or delete more sequence, but to make alignment increasingly fragile by rearranging existing blocks.

MAFFT diagnostics confirmed that the gradient affected alignment quality. From B1 to B4, mean alignment expansion increased from 1.08 to 3.79 and the gap fraction from 0.074 to 0.734. At B5 and B6, the gap fraction decreased to 0.671 and 0.593; this is not a recovery of homology, but a more compact form of alignment collapse in which block correspondence is no longer consistently represented. These values are used as alignment diagnostics rather than plotted in the main figure; full diagnostics, nRF values, and paired-test summaries are reported in SI Section S2.

\begin{figure}[!htp]
  \centering
  \includegraphics[width=1.0\linewidth]{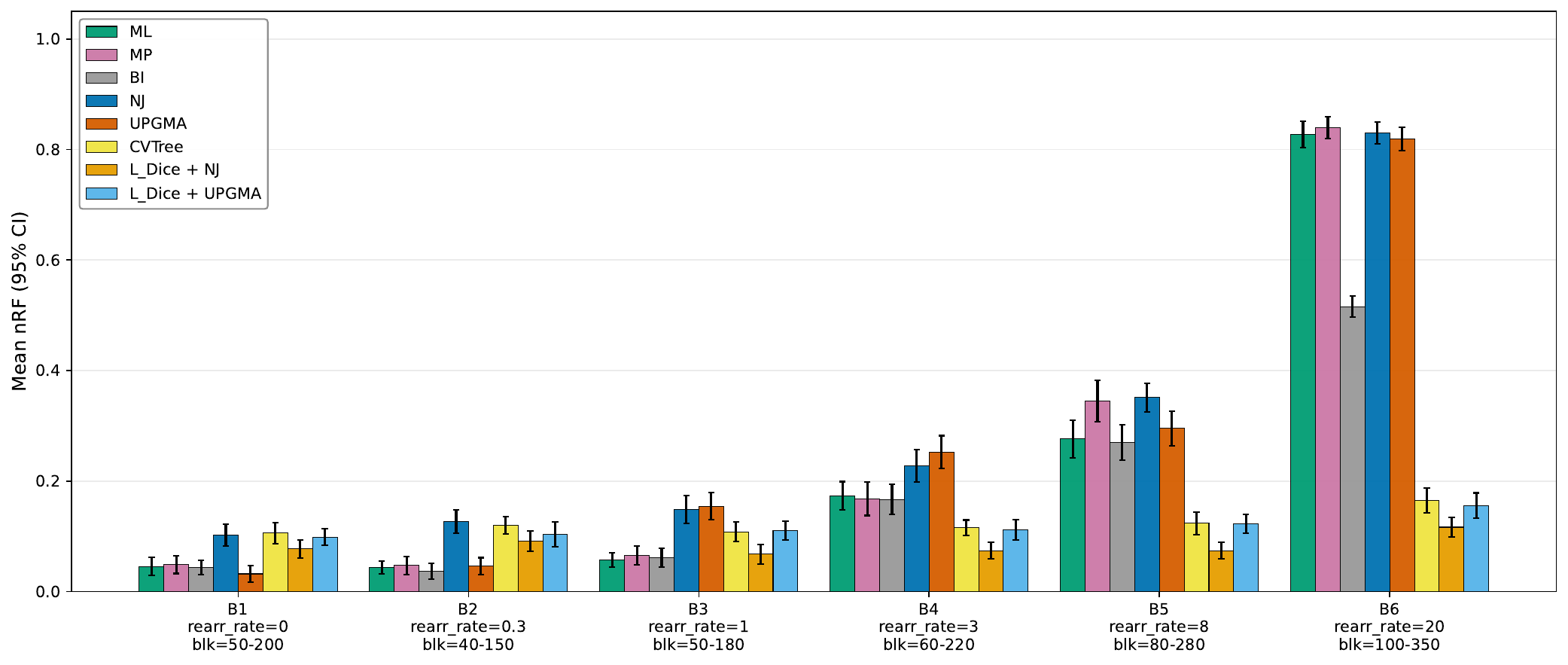}
  \caption{Block-translocation alignment-difficulty benchmark (B1--B6). As block rearrangement increases, alignment-free reconstructions remain comparatively stable, whereas MAFFT-dependent pipelines deteriorate after alignment begins to collapse. Bars show mean rooted nRF across 30 replicates; error bars are 95\% confidence intervals.}
  \label{fig:alignment-difficulty-trend}
\end{figure}

Figure~\ref{fig:alignment-difficulty-trend} summarizes the nRF trend. In B1 and B2, where block rearrangement is absent or rare, \LPD+NJ is accurate but not best: its mean nRF is 0.077 and 0.092, whereas UPGMA, MP, ML, and BI range from 0.032 to 0.049 in B1 and from 0.037 to 0.048 in B2. The selected paired tests show lower nRF for ML and BI in both conditions (largest Holm-adjusted $p=0.039$ in B1 and $p=6\times10^{-4}$ in B2), and the full table shows the same mean-level advantage for UPGMA and MP. This is the alignment-friendly part of the benchmark, where standard pipelines still have the advantage.

The pattern changes once rearrangement begins to damage the alignment. In B3, \LPD+NJ at 0.068 is significantly lower than classical NJ (0.149; $p=5\times10^{-4}$, Cliff's $\delta=+0.63$) and CVTree (0.108; $p=0.007$, $\delta=+0.57$); classical UPGMA (0.155) is also higher in mean nRF, while ML and BI remain comparable. From B4 onward, \LPD+NJ stays nearly flat (0.074--0.117), and CVTree shows a similar alignment-free plateau (0.115--0.165). The MAFFT-dependent pipelines deteriorate sharply. In B4, \LPD+NJ is lower than classical NJ, ML, and BI in the paired tests (Cliff's $\delta=+0.87$, $+0.80$, $+0.73$; $p\leq 2\times10^{-4}$), and also lower than classical UPGMA and MP in mean nRF (0.252 and 0.168). In B6, \LPD+NJ is 0.117, whereas classical NJ, UPGMA, MP, and ML range from 0.819 to 0.839, and BI is 0.515.

Thus, the useful point is not universal superiority. Under clean alignment, standard methods remain stronger. Under block rearrangement, the ladderpath approach loses less accuracy because it does not require the same positional homology representation.

%%%%%%%%%%%%%%%%%%%%%%%%%
\subsubsection{Structural signal under indel-dominated evolution}

The previous benchmark stressed alignment. The next one weakens substitutional signal and increases indel/duplication history. This asks whether phylogeny can be recovered from shared structural changes rather than mainly from aligned substitutions.

Substitutional divergence was kept weak and fixed throughout the benchmark. Five conditions (C1--C5) then increased indel/duplication intensity: the indel rate rises from 0.005 to 0.50, maximum indel length from 30 to 350 nt, and the indel-length distribution becomes progressively heavier-tailed. Each condition contains 30 paired replicates on shared 20-taxon Yule trees with 2{,}000 nt root sequences. Full design parameters, nRF values, and paired-test summaries are reported in SI Section S2. Unlike the block-translocation benchmark, this test is about whether shared indels and duplications carry recoverable tree signal when substitutions are sparse.

\begin{figure}[!htp]
  \centering
  \includegraphics[width=1.0\linewidth]{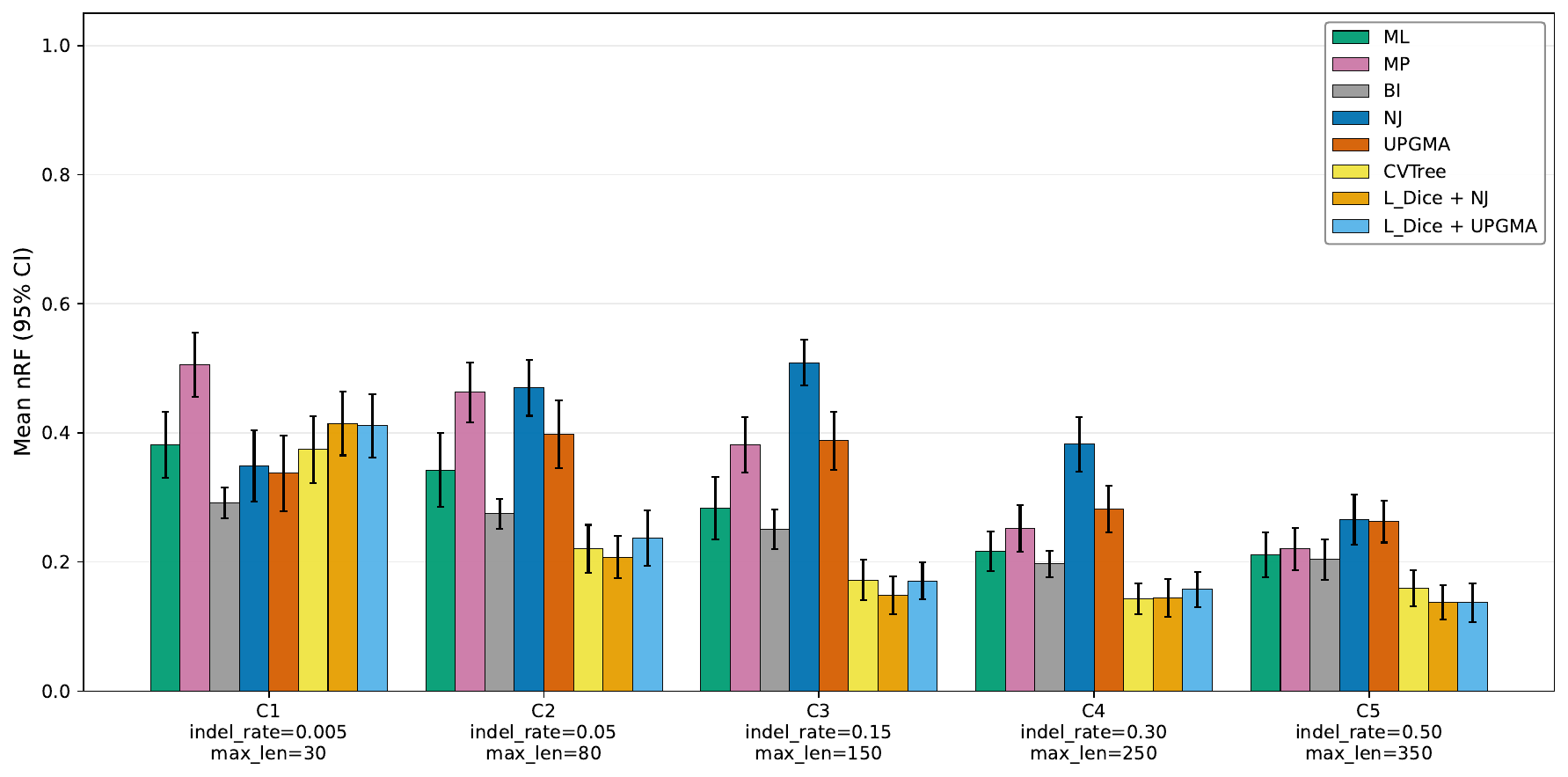}
  \caption{Indel-dominated structural-signal benchmark (C1--C5). Substitutional divergence is held weak; indel and motif-duplication intensity increase from C1 to C5. \LPD+NJ and CVTree remain usable across the gradient, whereas simple classical distance pipelines stay substantially less accurate. Bars show mean rooted nRF across 30 replicates; error bars are 95\% confidence intervals.}
  \label{fig:indel-dominated-trend}
\end{figure}

Figure~\ref{fig:indel-dominated-trend} summarizes the nRF trend. C1 is a signal-poor baseline, not a favorable case for the ladderpath approach: substitutions are weak, and indel/duplication structure is still minimal. \LPD+NJ has a mean nRF of 0.415; classical NJ (0.349) and ML (0.381) are slightly lower but not significantly so (Holm-adjusted $p=0.358$ and $p=0.171$), CVTree (0.374) is marginally lower ($p=0.045$), and BI is meaningfully lower at 0.292 ($p<10^{-4}$; mean difference $-0.123$). This condition is useful because the ladderpath approach should not be expected to recover signal when both substitutional and structural channels are weak.

As indel and duplication intensity increases, \LPD+NJ improves from 0.415 in C1 to 0.207, 0.148, 0.144, and 0.137 across C2--C5. CVTree follows a similar trajectory, from 0.374 to 0.159. Simple classical distance pipelines remain less stable: classical NJ improves from 0.349 in C1 to 0.266 in C5, but stays substantially above \LPD+NJ from C2 onward (C2: 0.470; C3: 0.509; C4: 0.382). The paired tests show a large gap from C2 onward (C2 mean difference $+0.262$, $p<10^{-4}$; C5 mean difference $+0.129$, $p=10^{-4}$).

ML and BI are more resilient than simple distance pipelines, but from C2 onward both remain above \LPD+NJ in mean nRF. In C2, \LPD+NJ is 0.207 compared with ML at 0.343 and BI at 0.275 ($p=6\times10^{-4}$ and $p=0.002$). In C5, \LPD+NJ is 0.137 compared with ML at 0.211 and BI at 0.204 ($p=0.009$ and $p=0.004$).

The result supports a specific claim: when substitutions are too sparse to anchor conventional inference, the ladderpath approach can use shared insertion, deletion, and duplication histories as phylogenetic signal. The claim remains bounded. CVTree is marginally lower than \LPD+NJ in C1 (mean difference $-0.041$, $p=0.045$) and statistically indistinguishable across C2--C5 (all Holm-adjusted comparisons non-significant), so the benchmark identifies a broader structural/compositional signal rather than a solution unique to Ladderpath.

%%%%%%%%%%%%%%%%%%%%%%%%%
%%%%%%%%%%%%%%%%%%%%%%%%%
\subsection{Representative empirical application scenarios}
\label{sec:empirical-scenarios}
The simulation benchmarks define the regimes in which Ladderpath is expected to be useful. We next use three empirical datasets as targeted checks rather than as a broad validation panel: a repeat-rich cpSSR marker \citep{toplin2008cyanidiales}, long structured organellar genomes from \textit{Musa} \citep{wu2021banana}, and the Fitch \& Margoliash cytochrome~$c$ protein dataset \citep{fitch1967}. These examples test different forms of sequence structure, so nRF is used here as a measure of topological congruence among methods or against a published reference, not as proof of absolute correctness.

%%%%%%%%%%%%%%%%%%%%%%%%%
\subsubsection{Repeat-rich empirical marker: cpSSR in Cyanidiales}

The Cyanidiales cpSSR dataset (\textit{Cyanidioschyzon merolae} IA-type strains from Yellowstone National Park) is a short repeat-rich marker. It contains 6 taxa (strains 5562, 5578, 5580, 5602, 5612, and 5633), each carrying ${\sim}5$ tandem copies of a ${\sim}136$\,bp repeat unit (total 675--678\,bp per strain; 678 columns once aligned). Such markers are widely used in plant ecology and evolution because they combine uniparental inheritance with high polymorphism \citep{provan2001, ebert2009}. Here the point is narrow: whether the ladderpath approach can give a coherent topology when repeat structure is a major part of the marker.

In this repeat-rich setting, \LPD+UPGMA, MP, NJ, and UPGMA converge on the same topology (nRF\,=\,0.000), recovering all three two-taxon clades \{5562,5633\}, \{5580,5602\}, and \{5578,5612\}. BI and ML each recover two of the three pairs but rearrange \{5580,5602\} (nRF\,=\,0.143 and 0.250 respectively; BI's lower value reflects a polytomy in the consensus). \LPD+NJ recovers \{5580,5602\} and \{5578,5612\} (nRF\,=\,0.250), and CVTree is the most divergent (nRF\,=\,0.750). The useful result is therefore not a general Ladderpath advantage, but a more specific one: a Ladderpath-Dice reconstruction can be fully consistent with several standard methods in a short repeat-rich empirical marker. Full pairwise nRF comparisons are provided in SI Section S3.

\begin{figure}[!htp]
  \centering
  \includegraphics[width=1.0\linewidth]{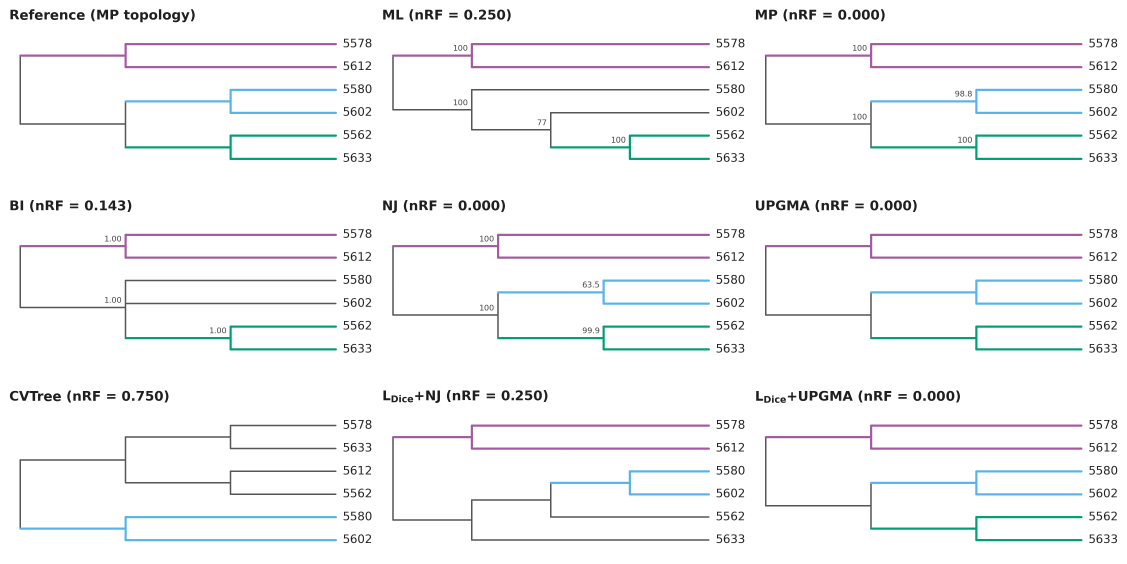}
  \caption{Tree comparison for the cpSSR dataset. Top-left: MP topology as reference (equivalent to NJ and UPGMA); other panels: inferred trees per method. Reference clades color-coded (\{5578,5612\} purple, \{5580,5602\} blue, \{5562,5633\} green); matching clades share the same color. Midpoint-rooted nRF vs MP topology annotated per panel. Branch labels: UFBoot (ML), non-parametric BS $B=1000$ (MP, NJ), PP$\times 100$ (BI).}
  \label{fig:cpssr-tree-comparison}
\end{figure}

\newpage
%%%%%%%%%%%%%%%%%%%%%%%%%
\subsubsection{Long structured organellar genomes: banana (\textit{Musa})}

The banana dataset tests a different boundary condition: long, structurally rich organellar genomes. We analyzed complete mitochondrial and plastome genomes from 21 Musaceae accessions \citep{wu2021banana}. The specific test is whether Ladderpath recovers the conflicting placement of a hybrid accession (\textit{Musa} $\times$ \textit{formobisiana}) in the mitochondrial versus plastome trees---a known consequence of differential organellar inheritance reported in the original study.

\LPD+NJ gives the closest Ladderpath agreement with the published ML reference tree of \citet{wu2021banana}, with nRF\,=\,0.189 for the mitochondrial genomes and 0.235 for the plastome genomes (Table~\ref{tab:banana-nrf}). \LPD+UPGMA is less close, with nRF\,=\,0.351 and 0.294, respectively. This table is not a full method-comparison panel; it reports the two main Ladderpath-Dice reconstructions against the published ML reference. The Ladderpath-derived trees also recover the main biological features reported by Wu et al., including the two-clade structure within \textit{Musa} and the contrasting placement of the hybrid \textit{Musa} $\times$ \textit{formobisiana} in mitochondrial versus plastome trees, consistent with organellar inheritance. Detailed tanglegram comparisons and corresponding \LPJ-based results are provided in SI Section S3.

Ladderpath therefore recovers both the major within-\textit{Musa} clade structure and the inheritance-driven discordance between mitochondrial and plastome trees directly from unaligned input. The alignment-free structural representation scales to genome-length organellar sequences while preserving the higher-order biological signal those genomes carry, indicating that sequence length itself is not a limiting factor of Ladderpath when sufficient structural organization is present.

\begin{table}[!htp]
\centering
\small
\begin{tabular}{lcc}
\toprule
      & Mitochondrial nRF to ref & Plastome nRF to ref \\
\midrule
\LPD+NJ & 0.189 & 0.235 \\
\LPD+UPGMA & 0.351 & 0.294 \\
\bottomrule
\end{tabular}
\vspace{0.4em}
\caption{Ladderpath-Dice nRF distances to the published ML reference tree for banana (\textit{Musa}) organellar genomes (21 taxa).}
\label{tab:banana-nrf}
\end{table}

%%%%%%%%%%%%%%%%%%%%%%%%%
\subsubsection{Deeply divergent protein sequences: cytochrome \textit{c}}

The third empirical case is the classic cytochrome~$c$ dataset of \citet{fitch1967}. It contains 20 taxa, from primates and ungulates to birds, reptiles, fish, insects, and fungi, with pairwise sequence identity ranging from $\sim$100\% (Human--Rhesus monkey) to $\sim$44\% (Human--\textit{Candida}). The proteins are short ($\sim$103--110 amino acids) and contain terminal indel variation. This makes the dataset a targeted check of whether Ladderpath remains informative on unaligned protein sequences across deep evolutionary splits.

In this dataset, MP and classical NJ tie at the best performance overall (nRF\,=\,0.111). \LPD+UPGMA reaches nRF\,=\,0.167, exactly matching classical UPGMA, and is lower than CVTree (0.222), ML (0.389), and BI (0.400). \LPD+NJ reaches 0.222, matching CVTree but above classical NJ. The corresponding \LPJ-based results give a comparable ranking, with \LPJ+UPGMA and \LPJ+NJ both at 0.167, so the Ladderpath variants in cytochrome~$c$ are tied with or match their classical analogues rather than fall below them. ML and BI are the two worst performers on this short, deeply divergent protein dataset. Full method-by-method comparisons are provided in SI Section S3.

\begin{table}[!htp]
\centering
\small
\begin{tabular}{llc}
\toprule
Method & Type &  nRF \\
\midrule
ML & Alignment-based  & 0.389 \\
MP & Alignment-based  & 0.111 \\
BI  & Alignment-based  & 0.400 \\
NJ & Alignment-based  & 0.111 \\
UPGMA & Alignment-based  & 0.167 \\
CVTree & Alignment-free  & 0.222 \\
\LPD+NJ & Ladderpath & 0.222 \\
\LPD+UPGMA & Ladderpath  & 0.167 \\
\bottomrule
\end{tabular}
\vspace{0.4em}
\caption{nRF distances to the Fitch (1967) reference tree for the 20-taxon cytochrome~$c$ dataset; lower values indicate closer agreement.}
\label{tab:cytc-nrf}
\end{table}

Ladderpath therefore recovers a biologically coherent deep evolutionary scaffold directly from unaligned protein sequences spanning roughly 44\% to 100\% pairwise identity. The Ladderpath trees agree with the 1967 reference on the major clade groupings---mammals, birds, fish, insects, and fungi---and the residual differences are concentrated in mid-level placements within the vertebrate part of the tree (SI Fig. S2), where cytochrome~$c$ is known to carry limited phylogenetic resolution. The short marker length and the distance-based historical reference set the scope of this test; within that scope the structural representation tracks the broad evolutionary architecture of the dataset.

%\newpage
%%%%%%%%%%%%%%%%%%%%%%%%%
%%%%%%%%%%%%%%%%%%%%%%%%%
\section{Discussion}

The present results support a broader theoretical point from complex-systems science: in molecular evolution, ``more is different,'' because whole-sequence architecture can carry information beyond independent site-wise changes \citep{anderson1972more,siegenfeld2020}. Standard phylogenetic models have been highly successful in modeling probabilistic change at homologous sites. Ladderpath complements this site-wise view by treating homology at a higher structural level: reusable sequence modules and their hierarchical relations. Drawing on algorithmic information theory, where related objects can be compared through shared algorithmic structure \citep{bennett1998information,li2001information,otu2003new,li2004similarity,cilibrasi2005clustering}, Ladderpath makes this shared structure explicit as hierarchical assemblies of reusable ladderons rather than as only aligned characters or scalar compression scores. The recovery of coherent phylogenetic signal when alignment is made difficult by indels, repeats, and sequence rearrangements therefore suggests that higher-order homologous architecture is not merely a descriptive property of biological sequences; it can function as an analyzable source of evolutionary information.

The simulation and anchor evidence describes a coherent pattern across evolutionary regimes. On the T7 bacteriophage anchor with a known experimental lineage, \LPD+UPGMA exactly recovers the reference topology, confirming that the structural distance reads phylogenetic signal correctly when the truth is independently known. Under standard substitution-driven simulations, Ladderpath is comparable to alignment-based pipelines once divergence is no longer extremely shallow. The pattern then diverges in the stress regimes: when alignment is destabilized by block translocations, \LPD+NJ remains topologically stable while ML, BI, MP, and classical NJ/UPGMA deteriorate together with their underlying alignment; and when sequence change is dominated by shared insertions, deletions, and duplications, \LPD+NJ improves further as more structural history becomes available for the decomposition to exploit. Taken together, these benchmarks position Ladderpath as a structural complement that becomes most useful precisely where alignment-based pipelines lose signal, while ML and BI remain the appropriate default when alignment is reliable and divergence is largely substitutional.

The empirical examples are consistent with this interpretation. In the repeat-rich cpSSR marker, one Ladderpath-Dice reconstruction is fully congruent with MP, NJ, and UPGMA, showing that repeat structure can yield a coherent topology rather than an idiosyncratic one. In banana organellar genomes, \LPD+NJ is the closest Ladderpath variant to the published ML reference in both mitochondrial and plastome datasets and recovers the expected contrast between organellar histories. In cytochrome~$c$, \LPD+UPGMA matches classical UPGMA at nRF 0.167, while MP and classical NJ are tied at the lowest score (0.111); Ladderpath thus tracks the simpler distance methods on this short, deeply divergent dataset without surpassing them. These examples do not establish a general empirical advantage, but they show that the structural signal highlighted by the simulations can appear in real datasets. As a bridge beyond topology reconstruction, the Brassicaceae collinearity analysis in SI Section S4 suggests that ladderons may also preserve interpretable genome-scale organization.

A further point is that ``alignment-free'' is not a single methodological category. CVTree, the alignment-free comparator closest to Ladderpath in design, is often near \LPD+NJ but does not reproduce the full \LPD+NJ advantage in the stress regimes examined here, indicating that fixed-length $k$-mer composition and nested structural reuse capture different aspects of sequence history. The paired-test summaries and supplementary nRF tables (SI Sections S2 and S3) support the same interpretation. Future work should therefore focus on faster implementations of the ladderpath decomposition, integration with model-based tree-search workflows, and broader empirical benchmarks in systems where structural evolution is expected to matter. Beyond tree reconstruction, we are also continuing to develop Ladderpath for genome-scale collinearity analysis and for the functional interpretation of long conserved ladderons; preliminary results on Brassicaceae syntenic blocks and the associated GO enrichment are reported in SI Section S4.

%%%%%%%%%%%%%%%%%%%%%%%%%
%%%%%%%%%%%%%%%%%%%%%%%%%
\section{Methods}

\subsection{Method panel and comparators}

The panel comprises maximum likelihood (ML), maximum parsimony (MP) \citep{fitch1971parsimony}, Bayesian inference (BI), neighbor joining (NJ/BIONJ) \citep{saitou1987nj,gascuel1997bionj}, UPGMA \citep{sokal1958upgma}, composition vector trees (CVTree) \citep{zuo2015}, and four Ladderpath variants (\LPD or \LPJ, each with NJ or UPGMA). Because Ladderpath returns a distance matrix rather than a tree, NJ and UPGMA play two roles: standalone methods on substitution distances, and tree-builders for Ladderpath. Classical NJ and UPGMA therefore serve as matched baselines for \LPD+NJ and \LPD+UPGMA---holding the tree-builder constant so that any topological difference isolates the contribution of the distance matrix. CVTree is the closest comparator in design: both are alignment-free and reduce sequences to substring features before distance-based tree building, but CVTree represents each sequence as a composition vector of fixed-length $k$-mer frequencies (with a Markov-model background subtraction), whereas Ladderpath uses nested, variable-length reusable modules.

\subsection{Sequence alignment and tree estimation}

Alignment-based methods used MAFFT v7.526 alignments \citep{Katoh2013}; the alignment-free methods (CVTree, Ladderpath) used raw sequences. ML used IQ-TREE 3.1.1 (empirical) or IQ-TREE 2.4.0 (simulations) \citep{Nguyen2015,Kalyaanamoorthy2017,Minh2020,Wong2026}, BI used MrBayes 3.2.7a \citep{Ronquist2012}, and MP used a Biopython v1.87 \citep{cock2009biopython} nearest-neighbor-interchange (NNI) heuristic (empirical) or the IQ-TREE 2.4.0 parsimony heuristic \citep{Minh2020} (simulations). NJ and UPGMA used DendroPy v5.0.8 \citep{sukumaran2010dendropy,moreno2024dendropy5} on the IQ-TREE ML distance matrix, and CVTree used an in-house composition-vector cosine distance \citep{zuo2015} ($k=6$ DNA, $k=3$ AA) via DendroPy NJ.

The representative tree per method was: the IQ-TREE UFBoot consensus (.contree) for empirical ML and the optimized .treefile for simulation ML; the MrBayes majority-rule consensus (25\% burn-in) for BI; the best-found NNI tree for MP; and the direct tree-builder output for NJ, UPGMA, and CVTree. For Ladderpath, CPU runs averaged the distance matrix over $K$ independent runs ($K=100$ for the empirical datasets, $K=10$ for the block-translocation and indel-dominated simulations, and $K=3$ for the standard-divergence simulation); the toy 12-taxon protein and banana organellar datasets used a single deterministic GPU run.

\subsection{Ladderpath distance computation}

Ladderpath analyses used the public code at \url{https://github.com/yuernestliu/lp_phylo} (ladderpath.py v2.0.18, lambda\_from\_laddergraph.py v2.1.0) \citep{liu2022,Liu2024npj,Liu2024PepHiRe,liu2024PRR,Liu2025Alg}, a greedy approximation; \(\lambda\) denotes the value returned by this fixed code path. Pairwise distances were computed as \(\lambda'(X)=\lambda(X)-1\) and \(\lambda'(X,Y)=\lambda(X,Y)-2\), yielding \LPD, with \LPJ the monotone Jaccard transform. Optional randomized \(\eta\)-estimation was disabled, and gaps and whitespace were stripped from any aligned input. The CPU implementation contains a wall-clock-based recompute trigger in find\_components (typical CV\,$\leq$\,0.5\% across replicate calls); the suffix-array GPU path (ladderpath\_gpu.py) is deterministic. All runs used the CPU implementation except the banana mitochondrial and plastome datasets and the toy 12-taxon protein dataset, which used the GPU path.

\subsection{Topological agreement (nRF)}

Topological agreement was scored by the normalized Robinson--Foulds distance (nRF) \citep{robinson1981comparison}: the rooted RF distance divided by $2(n-2)$, computed with ete3 v3.1.3 \citep{huerta2016ete3} (robinson\_foulds, unrooted\_trees=False). Each method tree was re-rooted on the same outgroup as its reference: the designated biological outgroup or midpoint for empirical analyses, and the smaller side of the reference tree's root split (anchored on its alphabetically first leaf) for simulations.

%%%%%%%%%%%%%%%%%%%%%%
\subsection{Toy-model tree comparison}

The toy-model analysis used 12 protein taxa (labeled 0--11; raw sequence lengths 12--39 residues, 39 amino-acid columns after MAFFT alignment). The binary reference tree used for nRF scoring (top-left panel of Figure~\ref{fig:toy-comparison}) is a discretization of the designed construction scheme (Figure~\ref{fig:toy-ground-truth}); all trees were rerooted on the designated outgroup before computing rooted nRF. The toy-model input sequences, reference topology, and analysis outputs are summarized in SI Section S1 and are provided in the public GitHub repository.

%%%%%%%%%%%%%%%%%%%%%%
\subsection{Simulation benchmarks}
\label{sec:random-seeds}

Simulated trees and sequences were generated with in-house Python 3.11.15 \citep{vanrossum2009python} scripts: a reference tree, GTR+\(\Gamma\) nucleotide evolution, then analysis-specific structural operations. Each condition had 30 replicates.

Randomness came from simulation generation and stochastic IQ-TREE/MrBayes searches; all other steps were deterministic for fixed inputs. Block-translocation used BASE\_SEED = 20260510: replicate $r$ tree seed $= BASE\_SEED+r$; condition $c$, replicate $r$ simulation seed $= BASE\_SEED\times100+ord(c_{\mathrm{last}})\times1000+r$; LBA perturbations added 999{,}983. Standard and indel-dominated analyses were deterministic per replicate without a numeric seed manifest; each replicate directory retains the realized tree and data, and per-replicate IQ-TREE2 and MrBayes seeds are recorded in run logs.

\subsubsection{Standard conditions}

Standard conditions: 20 taxa, ${\sim}4{,}000$-nt root sequences, independent random trees per replicate at tree scales 0.02, 0.06, 0.15, 0.30, 0.50. GTR+\(\Gamma\) with $\alpha=1.0$, light short indels, eight SSR loci per replicate (3--6-nt motifs; 8--15 repeat units); mean pairwise p-distances ${\sim}0.027$--$0.278$. MAFFT diagnostics: mean gap fraction $\leq 0.081$; mean alignment length ${\sim}4{,}016$--$4{,}360$ nt.

\subsubsection{Block translocation}

Block translocation: 30 taxa, 1{,}500-nt root sequences, paired Yule trees shared across the six conditions within each replicate. Substitution: GTR+\(\Gamma\), equal base frequencies, $\alpha=1.0$, relative rates AC$=$0.4, AG$=$1.0, AT$=$0.5, CG$=$0.5, CT$=$6.0, GT$=$0.4. Indels fixed across conditions (rate 0.02 per branch per nt; max length 6 nt; power-law $a=2.5$). Block-translocation Poisson rates 0.0, 0.3, 1.0, 3.0, 8.0, 20.0 events per branch; block sizes 50--200, 40--150, 50--180, 60--220, 80--280, 100--350 nt. Inversions and duplications excluded. MAFFT diagnostics: mean expansion $1.08\to3.79$; mean gap fraction $0.074\to0.734$ from B1 to B6.

\subsubsection{Indel-dominated evolution}

Indel-dominated: 20 taxa, 2{,}000-nt root sequences, paired Yule trees across C1--C5 per replicate, weak substitution (sub\_scale=0.01; $\alpha=0.5$). Indel rates C1--C5: 0.005, 0.05, 0.15, 0.30, 0.50 per branch per nt; max indel lengths 30, 80, 150, 250, 350 nt; power-law parameters 1.5, 1.4, 1.3, 1.2, 1.15. Clade-biased motif duplication probability 0.6. MAFFT mean gap fraction $\sim0.063$ (C1) $\to 0.634$ (C5).

%%%%%%%%%%%%%%%%%%%%%%
\subsection{Empirical application}

Empirical datasets: Hillis T7 bacteriophage lineage \citep{hillis1992, hillis1994}, Cyanidiales cpSSR (SSR9 variant 2 \citep{toplin2008cyanidiales}), banana mitochondrial and plastome genomes \citep{wu2021banana}, Fitch--Margoliash cytochrome~\(c\) protein \citep{fitch1967}, and a Brassicaceae collinearity dataset (SI Sections S1 and S4). Dataset sources, preprocessing notes, and repository organization are summarized in SI Section S1. Repeat-rich regions were not masked. Without a known history, nRF measured congruence among methods or against published reference topologies.

Dataset-specific ML and BI models, seeds, and BI diagnostics are in SI Table S2. Bootstrap and tree-level uncertainty are described in Section~\ref{sec:bootstrap-support}.

%%%%%%%%%%%%%%%%%%%%%%
\subsection{Support and statistical analyses}
\label{sec:bootstrap-support}

Bootstrap and tree-level uncertainty were computed for the revised empirical analyses (T7, cpSSR, cytochrome~$c$, toy 12-taxon protein). ML used IQ-TREE 3.1.1 with UFBoot \citep{Hoang2018UFBoot} and SH-aLRT \citep{Guindon2010SHaLRT}, $B=1000$ each; the .contree carries both values, UFBoot is plotted. MP used a non-parametric column-resampling bootstrap ($B=1000$) \citep{felsenstein1985}: the NNI heuristic was repeated on each resampled alignment from a JC/identity-NJ starting tree. NJ used the same scheme: each resampled alignment was passed to IQ-TREE 3.1.1 for an ML distance matrix and to DendroPy BIONJ. BI used MrBayes posterior probabilities from the majority-rule consensus. UPGMA, CVTree, and the LP variants are reported as topology only: UPGMA bootstrap is rarely used (ultrametric assumption); CVTree column-resampling is non-standard for an alignment-free method; LP support is the $K=100$ mean-distance standard error per species pair (typical SE $\leq 5\times 10^{-4}$). Internal node labels show UFBoot, non-parametric BS, and PP$\times 100$ as integers in $[0,100]$; nodes carrying only the default support of 1 are omitted.

Simulation results are means and standard deviations across 30 replicates. Within each condition-replicate cell, \LPD+NJ was compared with each other method by paired Wilcoxon signed-rank tests, reporting paired mean difference, 95\% CI, Cliff's $\delta$, and Holm-corrected $p$-values; positive differences indicate lower nRF for \LPD+NJ. $\alpha=0.05$ after Holm correction; full tables are in SI Section S2.1.

%%%%%%%%%%%%%%%%%%%%%%%%%
%%%%%%%%%%%%%%%%%%%%%%%%%
\section*{Conflict of Interest}
The authors declare that they have no conflict of interest.

\section*{Acknowledgements}
This study was funded by the National Natural Science Foundation of China (Grant No. 12205012 to Y.L.) and the Basic and Applied Basic Research Foundation of Guangdong Province (Grant No. 2025A1515012923 to Y.L.). Additional support was provided by a charitable donation from the China Resources Trust Gewu Basic Science Research Promotion Charitable Trust.

\section*{Data and Code Availability}
All data and code used in this study are publicly available at \url{https://github.com/yuernestliu/lp_phylo}.
\bibliographystyle{apalike}
\bibliography{references}

\end{document}